\documentclass[12pt,preprint]{aastex}
\usepackage{emulateapj5}
\shorttitle{Metallicity/LMXB No. density correlations}
\shortauthors{Maccarone, Kundu \& Zepf}

\begin{document}
\def\simlt{\mathrel{\rlap{\lower 3pt\hbox{$\sim$}}
        \raise 2.0pt\hbox{$<$}}}
\def\simgt{\mathrel{\rlap{\lower 3pt\hbox{$\sim$}}
        \raise 2.0pt\hbox{$>$}}}
\input epsf


\title{An explanation for metallicity effects on X-ray Binary properties}


\author{Thomas J. Maccarone\altaffilmark{1}, Arunav Kundu\altaffilmark{2} and Stephen E. Zepf\altaffilmark{2}}

\altaffiltext{1}{Astronomical Institute ``Anton Pannekoek,'' University of 
Amsterdam, Kruislaan 403, 1098 SJ, Amsterdam, The Netherlands; email: 
tjm@science.uva.nl}
\altaffiltext{2}{Department of Physics and Astronomy, Michigan State
University, East Lansing MI, 48824; email: akundu, zepf@pa.msu.edu}

\begin{abstract}

We show that irradiation induced stellar winds can explain two
important metallicity effects in X-ray binaries - the higher numbers
and the softer spectra of the X-ray binaries in metal rich globular
clusters compared to the metal poor ones.  As has been previously
noted by Iben, Tutukov and Fedorova, the winds should be stronger at
lower metallicity due to less efficient line cooling.  This will speed
up the evolution of the LMXBs in metal poor clusters, hence reducing
their numbers.  These winds can also provide extra material near the
accreting object which may create an intrinsic absorber to harden the
X-ray spectra of the metal poor cluster systems relative to the metal
rich ones, as suggested by observations.  We outline some additional
observational predictions of the model.

\end{abstract}

\section{Introduction}

Globular clusters (GCs) are important laboratories for studying
stellar populations, both main sequence and exotic.  X-ray binaries
are about 100 times overabundant in GCs compared to in the field
compared to the field (see e.g. the compilation of X-ray binary
properties in Liu, van Paradijs \& van den Heuvel 2001) due to
dynamical effects such as tidal capture and/or three and four body
interactions (e.g. Clark 1975; Fabian, Pringle \& Rees 1975; Hills
1976).  Furthermore, since detailed population studies of X-ray
binaries require more sources than the $\sim100$ in the Local Group,
the extragalactic case must be studied.  Given that the ages and
metallicities of individual extragalactic field stars are nearly
impossible to determine, but that these properties can be inferred
from integrated light of GCs, the GC X-ray binaries in other galaxies
represent an ideal place to study how age and metallicity affect X-ray
binary populations.

Clearly, the dynamical properties of a GC are most important for
predicting whether it will have an X-ray binary (e.g. Pooley et
al. 2003; Heinke et al. 2003). In addition, there also exists a strong
residual correlation between the number densities of X-ray binaries
and the metallicities of the GCs.  This was first suggested to be the
case in the Local Group (Grindlay 1993; Bellazzini et al. 1995), and
was shown more conclusively and dramatically in NGC~4472 (Kundu,
Maccarone \& Zepf 2002 - KMZ02).  The possibility that this represents
an age effect rather than a metallicity effect has been tested and
found not to be the case using data on NGC~3115 and NGC~4365 (Kundu et
al. 2003 - K03).  In NGC 3115, where the age spread is small, but the
metallicity spread is large, the metallicity effect was found to be at
least as strong as that in other early type galaxies.  Conversely, in
NGC~4365, where the age spread is large, the fraction of globular
clusters with LMXBs did not vary much with age and was similar to that
of older GCs with similar metallicities in other galaxies.  It thus
seems most likely that the LMXB population enhancement effects first
associated with the metallicity are, in fact, related directly to the
metallicity and not to some other correlated property such as age,
half-light radius, or distance from the center of the galaxy.  It has
also been noted that the metal poor Large Magellanic Cloud has a lower
ratio of LMXBs to HMXBs than the more metal rich Milky Way (Cowley
1994; Iben, Tutukov \& Fedorova 1997 - ITF97), which may be a
combination of the metallicity effects and the difference in their
star formation histories.

An additional metallicity effect can be found in the differences of
the soft X-ray spectra of these sources.  It has been found that the
spectra of the blue (i.e. metal poor) GCs are harder than those of the
red (i.e. metal rich) GCs in the Milky Way and M31 (Irwin \& Bregman
1998 - IB) and in NGC 4472 (Maccarone, Kundu \& Zepf 2003 - MKZ03).
On the other hand, the quiescent LMXBs seem to show no evidence of a
metallicity effect on the spectrum (e.g. Heinke et al. 2003), and the
effect becomes much weaker or disappears at higher X-ray energies
(Trinchieri et al. 1999; Di Stefano et al. 2002; Sidoli et al. 2001;
MKZ03).  

In this paper, we will show that the current theoretical explanations
for the overabundance effect are unlikely to match the typical factor
of 3 difference between the probabilities of metal poor and metal rich
GCs' having X-ray LMXBs.  We will put forth a new scenario invoking
the effects of irradiation induced winds (IIWs) can explain the
population difference.  We will also show that IIWs can explain the
previously unexplained spectral difference effects as a result of the
same physical process, with the same parameter values.

\section{Past Theoretical Work}
\subsection{IMF variations?}
Several attempts have been made to explain why metal rich GCs have
more LMXBs, but as yet, none seems satisfactory.  Grindlay (1993)
suggested that the correlation might be due to a flatter initial mass
function in higher metallicity GCs.  The general consensus seems to be
that the IMF is fairly universal (e.g. Kroupa 2002), and star
formation theory suggests that any metallicity dependence is likely to
be such that metal rich stars have typically lower masses since the
Jeans mass will be lower for the more efficiently cooling metal rich
gas (Larson 1998).  Although the present day mass functions of metal
rich Galactic GCs are flatter than those of metal poor Galactic GCs
(McClure et al. 1986), this seems to be largely due to the fact that
the metal rich systems are located more centrally within the Milky Way
and hence experience more extreme dynamical stripping of their low
mass stars, as seen observationally (Piotto \& Zoccali 1999) and
expected theoretically (Vesperini \& Heggie 1997; Baumgardt \& Makino
2003).

\subsection{Effects of stellar radius}
Later, Bellazzini et al. (1995) suggested that the larger stellar
radii of the stars in metal rich globular clusters might contribute to
their increased number densities of X-ray binaries.  Larger stars
should have a higher cross section for tidal captures and should
overflow their Roche lobes at larger separations.  The physics of
tidal capture are a hotly contested issue and some workers have
suggested that tidal captures have great difficulty in forming systems
similar to the X-ray binaries we observe (see e.g. Rasio \& Shapiro
1991; Rasio, Pfahl \& Rappaport 2000 and references within), while
others have suggested that the bulk of recycled pulsars may have been
formed in tidal capture systems (e.g. Di Stefano \& Rappaport 1992).
The alternative to formation of X-ray binaries by tidal captures is
formation through exchange interactions.  These are most commonly
three-body interactions where a neutron star encounters a binary
system composed of two non-compact stars and replaces one of those two
stars and forms a binary system with the other (see e.g. Clark 1975;
Hills 1976), or four-body interactions, where the neutron star is
already in a binary system when it interacts with another binary
system (see e.g. Mikkola 1984; see also Fregeau et al. 2003 for a
discussion of recent numerical work on 3-and 4-body interactions).

The easier Roche lobe overflow can be calculated using the standard
assumption of a binary separation distribution which is logarithmic,
i.e.  $P(a)\propto$$1/a$.  Taking Kepler's law and the period-mass
relation [equation (4.9) of Frank, King \& Raine (1992 - FKR92)] we
find that for a neutron star accretor and a typical (i.e. a 0.6
$M_\odot$) main sequence GC star donor, the Roche lobe overflow
orbital separation will be about 3 stellar radii, and varies little
with the donor star's mass.  The separation at which Roche lobe
overflow will occur increases linearly with the stellar radius for a
fixed stellar mass.  Given the logarithmic distribution of separations
and a metallicity dependence of radius such that
$R_{*}\propto$$Z^{1/8}$ (estimated from the stellar model
interpolation formulae of Tout et al. 1996), the contribution of the
term related to the increase in the number of Roche lobe overflowing
systems will be $1+\frac{1}{8} {\rm log}(\frac{Z_r}{Z_p})/{\rm log}
(\frac{r_{max}}{r_{min}})$, which is about 1.3 for a metallicity ratio
of 10.  Numerical calculations suggest that the tidal capture rate
goes as ${R_{*}}^{0.93},$ (e.g. Lee \& Ostriker 1986), so the number
of neutron star binaries formed by tidal capture will go as $Z^{0.12}$
which gives another factor of about 1.3 for a ratio of metallicities
of 10.  An additional small increase in the tidal capture probability
might come from the higher turnoff masses for the stars in metal rich
GCs (see e.g. Chaboyer et al. 1996), but this factor should be no more
than about 5-10\%.  Stellar radius effects thus seem unlikely to make
more than a $\sim$60\% difference (i.e. the product of the two 30\%
effects) in X-ray binary populations as a function of metallicity.  In
fact, since tidal captures are not likely to produce binary systems
that are similar to the observed parameters, it seems more likely that
the bulk of the LMXBs are formed in three and four body exchanges (see
e.g.  Rasio et al. 2000), and so the overabundance factor in red
clusters due to larger stellar radii is likely to be closer to the
factor of 1.3 than 1.6.

\section{The irradiation induced wind model}

The mass donors in X-ray binaries can absorb and reprocess
luminosities comparable to their intrinsic luminosities.  The mass
donor may then be ``puffed up'' to larger radii (e.g. Tutukov \&
Yungelson 1980; Podsiadlowski 1991; Harpaz \& Rappaport 1991) and that
the extra kinetic energy added to the envelope may drive off an
evaporative wind with a velocity of order the escape velocity from the
stellar surface (Arons 1973; Ruderman et al. 1989; Tavani \& London
1993; Pfahl, Rappaport \& Podsiadlowski 2003).  A fraction ($\sim$
10\%) of the gas lost by the mass donor will be accreted by the
compact star in much the same way that more typical stellar winds are
accreted in HMXBs.  The lifetimes of LMXBs will also be accelerated by
the extra mass loss in the IIWs (Tavani 1991a).

In addition to driving the evolution of the system, the irradiation
induced wind will also leave a large amount of gas in the environment
of the X-ray binary.  Most high mass X-ray binaries show clear orbital
modulations of the X-ray flux, especially at soft X-ray energies, and
this is taken to be evidence of internal absorption by the material in
the stellar wind.  An IIW from a low mass star will likely have a
velocity much closer to the orbital velocities in the binary system,
so the orbital modulation will not necessarily be as strong as in the
HMXBs.  Still, the increased absorption will have an effect on the
spectrum.

Let us summarize a few key past results on irradiation induced winds.
It is generally found that these winds should be more important in low
mass X-ray binaries than in high mass X-ray binaries, since in low
mass X-ray binaries the amount of radiation absorbed by the mass donor
from the accretion flow may far exceed the nuclear energy generation
rate and may hence have a significant effect on the structure of the
donor star (see e.g. Tutukov \& Yungelson 1980; Podsiadlowski 1991;
ITF97); IIWs may affect HMXBs as well (see e.g. Day \& Stevens 1993),
but have been much less well studied and are far less likely to
dominate the overall mass transfer.  Strong coronal winds are likely
to result from IIWs and may be self-sustaining even if the mass donor
does not fill its Roche lobe (see e.g. Basko \& Sunyaev 1973; Arons
1973); in fact, ITF97 have found that self-sustaining winds can
produce large luminosities even if the mass donor fills only
$\sim80\%$ of its Roche lobe, in agreement with past results that a
mass donor need not fill its Roche lobe if it is sufficiently
irradiated (Tavani, Ruderman \& Shaham 1989).  Some initial Roche lobe
overflow is likely to be required in order to start accretion, but if
the irradiation induced winds cause the mass loss to be substantially
faster than what would be caused by orbital and stellar evolution,
then the star may cease to fill its Roche lobe even as mass loss and
accretion continue.  Metal rich stars can dissipate much of their
absorbed energy through line cooling, while metal poor stars dissipate
this energy primarily through IIWs.  Interpolating from the stellar
cooling rate tables of Sutherland \& Dopita (1993), the mass loss rate
due to the IIWs should scale as $Z^{-0.35}$; the system lifetime
should scale as $Z^{0.35}$ if the mass loss is dominated by these
winds and the lifetime is determined by the timescale for the mass
donor to lose all its mass (ITF97).

As a caveat, we note that the treatment of ITF97, is based in part on
the analytical irradiation treatment in Iben, Tutukov \& Yungelson
(1995); while the results do agree well with the numerical work of
Tavani \& London (1993) within the parameter space of their models,
but ITF97 extrapolate outside this range. A more sophisticated
numerical treatment may be in order for future work, but is clearly
outside the scope of this paper.

For a wind emitted at the escape velocity from the mass donor,
equation (4.35) of FKR92 shows that the fraction of the wind captured by
the compact object will go as:
\begin{equation}
\frac{\dot{M}}{-\dot{M}_w}=\frac{1}{4}(\frac{M_{CO}}{M_{d}})^2(\frac{R_{d}}{a})^2
\label{f}
\end{equation}
where $\dot{M}$ is the mass accretion rate of the compact object
$\dot{M}_{w}$ is the wind mass loss rate, $M_{CO}$ is compact object's
mass, $M_d$ and $R_d$ are the mass and radius of the donor star, and
$a$ is the orbital separation.  Given typical values of
$a=4\times10^{11}$ cm, $R_{d}=7\times10^{10}$ cm, $M_{d}=0.6 M_\odot$
and $M_{CO}=1.4 M_\odot$, about 5\% of the wind is accreted.

To simplify the calculation, we make one additional assumption, that
the red GCs accrete from a combination of Roche lobe overflow and
IIWs, while the bright LMXBs in blue GCs accrete from a
self-sustaining wind.  We will later revisit this assumption and show
that it is not necessary in order to reproduce roughly the
observations.  We now define the notation for the following 5
equations - $\dot{m}$ indicates the mass accretion rate onto the
compact object, while $\dot{M}$ is the mass loss rate.  The subscript
$w$ indicates wind mass loss, $RL$ indicates Roche lobe overflow
effects, $p$ indicates systems with metal poor donors, and $r$
indicates systems with metal rich donors, $N$ indicates the number of
systems and $Z$ indicates metallicity. Then
\begin{equation}
\dot{m} = \epsilon \dot{M}_{w} + \dot{M}_{RL},
\end{equation}
where $\epsilon$ is the fraction of the mass lost in the wind that is
accreted by the compact object and is given by equation (\ref{f}).
Then, starting from the assumption outlined above that the Roche lobe
overflow component is neglible for the metal poor systems, we have:
\begin{equation}
\dot{m}_{p}=\epsilon \dot{M}_{w,p},
\label{p}
\end{equation}
while for metal rich stars,
\begin{equation}
\dot{m}_{r}=\epsilon (\frac{Z_{r}}{Z_{p}})^{-0.35} \dot{M}_{w,p} + \dot{M}_{RL},
\label{red}
\end{equation}
where $\dot{M}_{RL}$ is the mass loss rate due to the Roche lobe
overflowing component of the accretion flow, and is assumed to be much
smaller than the irradiation driven mass loss rate for the case of a
metal poor donor.  Combining equations (\ref{p}) \& (\ref{red}), we find
that, for the same luminosity,
\begin{equation}
\frac{\dot{M}_{r}}{\dot{M}_{p}} =
(\frac{Z_{r}}{Z_{p}})^{-0.35} + \epsilon - \epsilon
(\frac{Z_{r}}{Z_{p}})^{-0.35}.
\label{summary}
\end{equation}
The fraction of the mass loss in the metal rich systems coming from
the IIW is $(\frac{Z_{r}}{Z_{p}})^{-0.35}$.

%It is important to note that because the luminosity of the sources in
%blue GCs will tend to be increased due to the higher
%mass loss rate in the irradiation induced stellar wind, so the effects
%of the luminosity function must be considered.  The luminosity
%functions of point sources in elliptical galaxies' globular clusters
%in the X-rays typically show broken power laws, with
%$N(>L)\propto$$L^{-\beta}$ below a cutoff and
%$N(>L)\propto$$L^{-\gamma}$ above a cutoff, and with the cutoff
%typically around $2\times10^{38}$ ergs/sec (KMZ02).  The cutoff
%luminosity is generally quite close to the Eddington limit for a
%neutron star, although it is far from certain whether this is the true
%reason for the cutoff.  As the bulk of the sources detected in the
%well-observed galaxies will be below the high luminosity cutoff, a
%reasonable simplification of the luminosity function is to treat it as
%a single power law.  Perhaps counterintuitively, given a flux-limited
%sample and a power law luminosity function $N(>L)\propto$$L^{-\beta}$,
%a systematic increase by a constant multiplicative factor $f$ in the
%luminosities of all sources in the population will {\it not} change
%the mean observed luminosity; instead, it will increase the observed
%number of sources by $f^{\beta}$.
The number of X-ray binaries should scale as the formation rate times
the lifetime.  The formation rate effects have been studied by
Bellazzini et al. (1995), and given their lines of argument, we found
in Section 2.2 that the stellar radius effects should produce a
difference by a factor of about 1.3 (if exchange interactions
dominate) to 1.6 (if tidal captures dominate).

The effects of IIWs are predominantly on the system lifetimes.  For
systems at a given luminosity, equation \ref{summary} shows the
difference in mass loss rate, the inverse of which gives the ratio of
source lifetimes.  Thus we find that
\begin{equation}
\frac{N_{r}}{N_{p}} = \frac{1}{(\frac{Z_{r}}{Z_{p}})^{-0.35}
+ \epsilon - \epsilon (\frac{Z_{r}}{Z_{p}})^{-0.35}}.
\label{finalratio}
\end{equation}
To compute the actual ratio of the number of red to blue GC X-ray
sources, it is necessary to multiply the factor of 1.3 to 1.6 from
the stellar radius effects by the value from equation
\ref{finalratio}, which should be about 2.1 for the typical parameters
$\epsilon=0.05$ and $\frac{Z_{r}}{Z_{p}}$=10. This gives a factor
between about 2.6 and 3.4, although this factor is likely a slight
overestimate, because even the X-ray sources in the most metal poor
GCs are likely to have at least some Roche lobe overflow contribution
to their X-ray luminosities.

It has been assumed that the stellar wind velocity is equal to the
escape velocity from the surface of the star; this need not be the
case.  The extra wind energy for the metal poor stars may be
dissipated as a higher velocity wind rather than as a more dense wind.
The fraction of the mass lost that is accreted scales as
$v_{wind}^{-4}$ (FKR92), which alternatively scales as the inverse of
the wind power squared, for a constant mass loss rate.  The luminosity
of the LMXBs in metal poor GCs would then be surpressed by a factor of
about $(\frac{Z_r}{Z_p})^{0.70}$, while the lifetimes of the two
classes of systems would be about the same.  Because the luminosity
function has a slope of about ${-0.55}$ (KMZ02), the ratio of the
number of metal rich and metal poor cluster X-ray systems would then
be $(\frac{Z_r}{Z_p})^{0.39}$, which, for the canonical factor of 10
difference between the two modes gives a factor of about 2.5
difference in the expected number of observed systems.  We do note
that the there might be systematic variations in the slope of the
luminosity functions as a function of metallicity, but absent
measurements or a theoretical model, we assume they will be the same.

We wish now to estimate the contribution to the column density with
which these systems will typically be observed due to the IIW.  We
find an average density of mass in a sphere around the mass donor with
radius equal to the diameter of the orbit.  For a path length of the orbital radius, the column density $N_H$ is then:
\begin{equation}
N_H=\frac{\dot{M}}{8v_{w}r_{orb}\mu},
\end{equation}
where $\mu$ is the mean molecular weight of the gas in the wind,
$v_{w}$ is the wind velocity, and $r_{orb}$ is the orbital separation.
This value, $\approx6\times10^{21}$cm$^{-2}$, should give a good
approximation over orbital phase and inclination angle for the column
density observed.  Edge-on sources might be expected to have higher
column densities, but the inclination angles for the GC LMXBs are not
well constrained.

\section{Observational evidence for the scenario}
\subsection{NGC 4472 - number of sources}
NGC~4472 is the first galaxy where the metal rich mode was shown to
have a higher fraction of GCs with LMXBs than the metal poor mode
(KMZ02).  It seems unlikely that the GCs in NGC 4472 span a wide range
of ages; they are all likely to be within a factor of 1.5 to 2 in age
(Beasley et al. 2000; Cohen, Blakeslee \& C\^{o}t\'{e} 2003).  The
metallicity effects on the number of globular cluster X-ray sources
thus are not likely due to an age difference between the metal rich
and metal poor GCs.  The age measurements are rather sensitive to the
stellar populations models for the Balmer lines and there could still
be a rather substantial age difference between the two samples.  More
strict constraints are the age measurements of Puzia et al. (2002)
which confirm that the correlation between metallicity and LMXB
specific frequency in NGC 3115 is due to metallicity and not age
(K03), but this system has fewer X-ray sources, so the ratio of the
number of LMXBs in metal rich and poor clusters cannot be as well
determined.  Finally there is the case of NGC 4365, where the ages do
span a rather wide range, but do not seem to be strongly correlated
with the LMXB number density (K03).

The two modes in color for NGC 4472 peak at $V-I$ of 0.98 and 1.23
(KMZ02), corresponding to values of [Fe/H] of -1.26 and -0.08,
respectively, according to the scaling law of Kundu \& Whitmore
(1998).  Defining the metal rich/metal poor mode boundary to be
$V-I$=1.10, we find that 23 of the 450 metal rich GCs and only 7 of
the 370 metal poor GCs contain X-ray sources.  The metal rich GCs
are thus $2.7\pm1.2$ times as likely to contain X-ray sources as the
metal poor GCs.
\subsection{NGC 4472 - source spectra}
The spectra also show a difference as a function of metallicity.
While the individual spectra cannot be easily measured because of the
low count rates, we have found that the summed spectra in NGC 4472 are
harder in the blue GCs than in the red ones (MKZ03).  If we hold the
neutral hydrogen column in both cases to the Galactic value of
1.6$\times10^{20}$ cm$^{-2}$ and fit a power law model to the data, we
find a spectral index of 1.02$\pm$.27 for the blue GCs while we find a
spectral index of 1.46$\pm$.10 for the red GCs (90\% error contours).
Allowing the column to float freely for the red GCs, we find $N_H$ to
be 4.8$\times10^{20}$ cm$^{-2}$ and the power law index to be 1.57.
Then, we fix the power law index for the blue GCs and find that the
data is best fit with a column density of 1.1$\times10^{21}$
cm$^{-2}$.

We note that the solution to the spectral difference problem is not
unique and is prone to numerous systematic uncertainties.  The
photoelectric absorption models we have used assume a solar
composition for the absorbing medium, and that the medium is cold
(i.e. completely un-ionized).  The underlying spectrum for accreting
neutron stars and black holes in the 0.5-8 keV range is unlikely to be
a single power law.  Finally, we have averaged over many systems with
different values of $N_H$ and with different underlying spectra.
Still, the rough information given, that the metal poor GCs have an
intrinsic absorption of about $10^{21}$ cm$^{-2}$ and that the column
density is about 3 times as large for the metal poor GCs as it is for
the metal rich GCs seems to be a reasonable inference to draw from the
data.  The theoretical model predicts a higher value for the blue
clusters, but the linear averaging tends to overestimate the effects
on the spectrum, so the fact that our crude calculation over-predicts
the amount of absorption is to be expected.  Furthermore, the fact
that the gas is likely to be partly ionized and will make the fitted
value of the $N_H$ less than the actual value, and also some of the
gas mass will condense into a geometrically thin accretion disk, and
hence will have an effect of absorbing X-rays only if the inclination
angle is very low.  That the values are on the same order of magnitude
and that the blue clusters have about 3 times as much absorption in
the fits is about as good an agreement as can be expected given the
crude modelling and the considerable theoretical uncertainties in the
models of IIWs.

\subsection{Local Group Sources}
Both the Milky Way and the Magellenic Clouds have been rather well
studied in terms of their X-ray stellar populations, and they, show a
metallicity difference on the same order as that between the metal
rich and metal poor modes for GCs - about a factor of 10.  As only a
small fraction of the X-ray sources in any of these galaxies is in a
GC, the star formation rate has a substantial effect on the relative
number densities of X-ray binaries, so merely comparing number counts
per unit stellar mass is not likely to prove frutiful.  However, one
can be fairly confident that the high mass X-ray binary population is
not heavily affected by IIWs because the luminosities of high mass
stars are much larger than the intercepted and absorbed luminosities.
Therefore, the ratio of LMXBs to HMXBs might give a rough estimate of
how important the IIWs are.  The suggestion of ITF97 that the
difference of this ratio might be indicating that irradiation induced
winds are playing an important role is therefore additional evidence
in favor of this scenario.

\section{Potential observational tests}

This model makes several testable predictions.  The first is that
there should be a monotonic dependence between the number of LMXBs per
unit stellar mass and the metallicity; given enough statistics we
should see a difference in the LMXB specific frequency as a function
of the metallicity itself, and not just as a function of whether a
cluster is in the metal rich or metal poor mode.  Much new data has
recently entered the Chandra archives, so it is now possible to test
this prediction.  There is already a correlation over a range of
metallicities in the ROSAT spectral indices of GC X-ray sources which
does not show a ``critical metallicity'' (IB), so our model seems to
pass this test so far.

This scenario also predicts that neutron star LMXBs will be affected
far more than other types of ``dynamically interesting'' sources.
Blue stragglers and cataclysmic variables will not generate high
enough X-ray luminosities to excite substantial IIWs.  Black hole
systems, because of the higher mass compact objects, will accrete the
IIW gas much more efficiently.

IIWs have been suggested to explain why systems such as 4U 1820-30
show different period evolution than would be expected from
conservative mass transfer driven by gravitational radition (Tavani
1991b).  An alternative is that the system is being effected by the
interactions with the GC potential (van der Klis et al. 1993).
Gravitational wave observations from future missions such as LISA may
help break this degeneracy.

Additionally, our scenario predicts that the metallicity effects
should be essentially the same for field sources as they are for
globular cluster sources.  Given two galaxies with similar star
formation histories, the more metal rich galaxy should have more field
X-ray binaries.  The natural way to test this hypothesis would be to
look at the field X-ray binary populations of elliptical galaxies, as
(1) they have very little recent star formation and (2) metallicity
tends to scale with galaxy mass.  A potential problem with this
approach is that a fraction (and indeed, perhaps a large fraction) of
the field X-ray binaries may have been created through stellar
interactions in GCs and released into the field through dynamical
ejections or through tidal destruction of the GCs (see MKZ03 and
references within; see also Grindlay 1988).  Given that both the tidal
destruction rate and the metallicity are likely to be correlated with
the mass, applying this test is not straightforward.  On the other
hand, the field sources of elliptical galaxies should show a
metallicity effect on their energy spectra regardless of concerns over
formation processes.

A better test might then be to extend the suggestion of ITF97 that the
difference in the ratio of LMXBs to HMXBs in the Milky Way and the LMC
is due to the metallicity difference.  Since most HMXBs are accretion
powered pulsars, a reasonably good separation between the bright ends
of the luminosity distributions of HMXBs and LMXBs should be possibly
given good Chandra spectra of nearby spiral galaxies.  We note that
this is a generic prediction of any model in which the metallicity
effects are strictly due to metallicity, but will provide a way to
distinguish between true metallicity effects and effects of
metallicity being correlated with more difficult to measure
parameters, such as the dynamics of the system.

Past work on detailed spectral fitting on Milky Way GCs has been
suggested to indicate that there is little evidence for intrinsic
absorption is these systems (Sidoli et al. 2001).  We note that this
is not in conflict with our model.  Sidoli et al (2001) did not obtain
a satisfctory spectral fit to the data for M 15, the most metal poor
of the Milky Way's globular clusters with an X-ray source, probably
because BeppoSax was not capable of resolving the two bright X-ray
sources in the cluster (White \& Angelini 2001).  The other two most
metal poor clusters, NGC~1851 and NGC~6712 do show excess absorption
in the X-rays compared with the optical, and the other globular
clusters in the Milky Way all have optical extinctions significantly
higher than the excess predicted by our model, so the fits would not
be very sensitive to intrinsic absorption.  Additionally, we note that
the fitting of Sidoli et al. (2001) was done using the standard
assumption that the absorber would be cold material of solar
composition.  Since BeppoSax is sensitive to absorption edges in the
$\sim$ few keV range, this may cause a systematic error in the fitted
absorption value, as noted above.  The results of Sidoli et al. (2001)
certainly place upper limits on the amount of intrinsic absorption in
the Milky Way's LMXBs, but these upper limits are mostly too high to
place strong constraints on our model.  A more sensitive test would
come from XMM-Newton spectra of M31 globular cluster X-ray sources,
where there will be little non-intrinsic absorption since the globular
clusters will not be viewed through the disk of the Galaxy.  In fact,
many of the M31 globular cluster X-ray sources show evidence for
intrinsic absorption, and the ones that do are predominantly in metal
poor clusters (Irwin \& Bregman 1999).

Differences in the luminosity functions between red and blue globular
cluster X-ray sources should also, in principle, provide a way to
discriminate between models for formation and evolution of their X-ray
binaries.  Unfortunately, it is difficult at this time to make a
prediction from our scenario.  It is not clear on theoretical grounds
whether the extra wind energy in metal poor systems manifests itself
as a higher mass loss rate, yielding probably slightly higher
luminosities, albeit for much shorter amounts of time, or as higher
wind velocities, in which case the efficiency of wind capture is
lower, so the luminosity will be lower at a given mass loss rate, or
as some combination of the two.  Furthermore, there is not yet a
sufficiently large sample of X-ray binaries in globular clusters for
making a good comparison of the luminosity functions.  This does
remain a good test to bear in mind for future work.

\section{Conclusions}

We have outlined a scenario whereby the two metallicity effects seen
in LMXBs in globular clusters, higher number density in metal rich
clusters, and harder low energy X-ray spectral in metal poor clusters,
can be explained via the same mechanism -- irradiation induced stellar
winds.  We have presented additional feasible observational tests of
this picture.  While we have shown that the physics of the irradiation
induced winds required to reproduce the observations is consistent
with the most recent theoretical work, we also note that this is a
rather complicated problem which is deserving of considerable
additional attention by experts in binary stellar structure and
evolution.  We hope this paper will help to stimulate such work in the
future.

\section{Acknowledgments}
We acknowledge funding from NASA LTSA grants NAG5-12975 and
NAG5-11319.  We thank Michael Sipior, Simon Portegies Zwart and
Jasinta Dewi for useful comments on this manuscript.  We thank the
referee, Josh Grindlay, for suggesting that we improve the clarity of
the paper and the depth of the background discussion.

\end{document}